\documentclass{elsart}

\usepackage{harvard}

\usepackage{epsfig}

\usepackage{amssymb}

\def\apj{ApJ}
\def\apjl{ApJL}
\def\aj{AJ}
\def\nat{Nature}
\def\mnras{MNRAS}

\def\url#1{{\ttfamily\def\/{/\discretionary{}{}{}}#1}}

\begin{document}

\begin{frontmatter}
\title{Long-Term Evolution in Transit Duration of Extrasolar
Planets from Magnetic Activity in their Parent Stars}
\author{Abraham Loeb}
\ead{aloeb@cfa.harvard.edu}
\address{Astronomy Department, Harvard University, 60
Garden St., Cambridge, MA 02138, USA}

\begin{abstract}
Existing upper limits on variations in the photospheric radius of the
Sun during the solar magnetic activity cycle are at a fractional
amplitude of $\sim 2\times 10^{-4}$. At that level, the transit
duration of a close-in planet around a Sun-like star could change by a
fraction of a second per year. This magnitude of variation is larger
than that caused by other studied effects (owing to proper motion or
general-relativistic effects), and should be included in the analysis
of constraints on multi-planet systems from transit timing. A temporal
correlation between the transit duration and spectroscopic measures of
stellar activity can be used to separate the stellar radius change
from other effects.  The magnetic activity effect could be
significantly larger for late-type stars, such as M-dwarfs, which are
more variable than the Sun.  In general, precision transit
measurements provide a new tool for measuring long-term variations of
stellar radii.

\end{abstract}

\end{frontmatter}


\section{Introduction}

Recently, it has been suggested that precise measurements of transit timing
variations can reveal the existence of additional planets in known
planetary systems (Miralda-Escud{\'e} 2002; Agol et al. 2005; Holman \&
Murray 2005; Heyl \& Gladman 2006). Related observations are already
placing interesting constraints on planetary companions (Steffen \& Agol
2005; Agol \& Steffen 2005; Ribas et al. 2008; Miller-Ricci et al. 2008a),
despite the intrinsic noise introduced by stellar activity (Miller-Ricci et
al. 2008b).  Space missions, such as
MOST\footnote{http://www.astro.ubc.ca/MOST/},
COROT\footnote{http://smsc.cnes.fr/COROT/} and
Kepler\footnote{http://kepler.nasa.gov/}, are capable of detecting timing
variations over their lifetime at a sensitivity better than a second per
year (P{\'a}l \& Kocsis 2008; Jord{\'a}n \& Bakos 2008; Rafikov 2008).

Aside from the dynamical interaction of the transiting planet or parent
star with another object, it was recognized that general relativistic
effects (P{\'a}l \& Kocsis 2008; Jord{\'a}n \& Bakos 2008) as well as the
proper motion of the system (Rafikov 2008), would introduce variations in
the transit timing and duration at levels smaller than a second per year.
Here, we point out that the magnetic activity cycle of the host star may
cause long-term variations in the stellar radius on timescales of years
that could make an even larger contribution to the observed transit
duration variations. In \S 2 we calibrate the magnitude of this effect
based on existing data for the Sun. Finally, in \S 3 we comment on the
implications of future transit measurements of these magnetic activity
variations in other stars.

\section{Method}

The transit duration of a planet on a circular orbit of radius $a$ around a
star of mass $M_\star$ and radius $R_\star$ is given by, 
\begin{equation}
t_{\rm tran}= 2{R_\star(1-p^2)^{1/2}\over (GM_\star/a)^{1/2}},
\end{equation}
where $p$ is the minimum separation between the planetary trajectory and
the stellar disk center on the sky, in units of $R_\star$.  The fractional
change in the transit duration timing (TDV) owing to a small fractional
variation in the stellar radius\footnote{A change in the image shape of the
parent star could change the transit timing in a more complicated
way. Limits on fractional variations in the oblateness of the Sun during
the solar cycle are at the level of $\sim 10^{-5}$ (Kuhn et al. 1998).},
$\Delta_{R_\star}$, is given for $\Delta_{R_\star}\ll [(1-p^2)/2p^2]$ by
\begin{eqnarray}
\nonumber &&\Delta t_{\rm tran}= \Delta_{R_\star}t_{\rm tran}=
\left(1-p^2\right)^{1/2} \times \\ && \times 1.49~{\rm s}
\left({\Delta_{R_\star} \over 10^{-4}}\right) \left({M_\star \over
M_\odot}\right)^{-1/2}\left({R_\star\over R_\odot}\right) \left({a\over
0.1{\rm AU}}\right)^{1/2},
\label{basic}
\end{eqnarray}
where $\Delta_{R_\star}\equiv \left(\Delta R_\star/R_\star\right)$.  For
$\Delta_{R_\star}\gg [(1-p^2)/2p^2]$, the transit duration could either
diminish as the stellar image shrinks or increase by $\Delta t_{\rm
tran}=[2p^2\Delta_{R_\star}/(1-p^2)]^{1/2}t_{\rm tran}$ as the image grows,
but we regard this grazing-incidence regime as rare and unlikely.

The solar magnetic activity cycle is known to cause fractional changes
of $\sim 0.1\%$ in the distribution of the near-UV, visible, and
near-IR brightness of the Sun relative to the ecliptic plane between
solar minimum and maximum (Fr{\"o}hlich 2006; Foukal \& Bernasconi
2008). At solar maximum, cold spots cover a larger fraction of the
surface area of the Sun, but the surface brightness in between the
spots increases due to energy deposition by magnetic activity.
Helioseismological data implies variations in the rotational kinetic
energy at a fractional level of $\sim 0.2\%$ (Antia, Chitre, \& Gough
2008), as well as in the sound speed at the base of the convective
zone at a fractional level of $(\Delta c^2/c^2)\sim 0.7\times 10^{-4}$
(Baldner \& Basu 2008), over the solar cycle. The associated change in
the pressure supporting the gas would inevitably produce a variation
in the physical radius of the star.  In the context of TDV, we are
interested in the variation of the apparent size of the photospheric
stellar disk on the sky, where the optical depth to emergent radiation
is of order unity.

State-of-the-art studies of the solar radius place an upper limit
$\Delta_{R_\odot}<2\times 10^{-4}$ on its fractional variation over
periods of decades (Djafer, Thuillier, \& Sofia 2008; Lefebvre et
al. 2006; Kuhn et al. 2004).  Reports on the detection of a variable
solar radius have been controversial (e.g., Noel 2004, Lefebvre \&
Kosovichev 2005).  Helioseismological data implies a $\sim 10^{-4}$
change in the sound speed at the base of the convective zone (Baldner
\& Basu 2008).  However, the fractional variation in the solar radius
could be much smaller (see also Stothers 2006). Nevertheless, it is
interesting to examine the sensitivity of future transit timing
measurements to variations of $\Delta_{R_\odot}\sim 10^{-4}$ over the
solar cycle.  Substituting this (upper limit) value in equation
(\ref{basic}) leads to a TDV at a level of $\sim 0.3 (a/0.1{\rm
AU})^{1/2}$ seconds per year over a cycle of $\sim 5.5$ years in
transits of close-in planets around Sun-like
stars\footnote{Shorter-term variations with a higher amplitude
(flares) introduce noise for TDV in all stellar
types.}. Interestingly, these variations are larger than other effects
which have been considered in the literature as detectable (P{\'a}l \&
Kocsis 2008; Jord{\'a}n \& Bakos 2008; Rafikov 2008), and so we
conclude that a search for TDV would provide interesting new limits on
stellar radius variations that are competitive with state-of-the-art
measurements for the Sun.  A generic signature of the TDV induced by a
circularly symmetric variation in stellar radius is that the timing of
the transit center remains unchanged as the transit duration changes.

\section{Discussion}

Precision timing of planetary transits offers a new tool for measuring
long-term variations in stellar radii. Such measurements would provide
unprecedented constraints on the surface conditions during the
magnetic activity cycles of stars at levels that are not accessible
for the Sun.  In difference from rare transits within the solar
system, the short orbital period of close-in planets around other
stars and the use of the cumulative flux from the stellar disk offer
improved prospects for averaging over fluctuations in the stellar limb
(owing to surface spots or other phenomena).

The magnetic activity origin of the TDV can be revealed through its
temporal cross-correlation with stellar brightness variations (in
analogy to the Sun; see Fr{\"o}lich 2006) and with spectral indicators
such as the Ca K line (Baliunas 2006).  A temporal correlation of this
type can, for example, help to distinguish between the effect of a
stellar radius change and a change in the inclination of the planet's
orbit.  In general, the magnetic cycle effect should be kept in mind
when inferring dynamical constraints on multi-planet systems or on
proper motion from transit timing.

We emphasize that the TDV induced by magnetic activity could be
significantly larger for late-type stars, such as M-dwarfs, which are more
variable than the Sun (Schmidt et al. 2007; Lane et al. 2007; Rockenfeller
et al.2006).

\bigskip
\noindent
{\bf Acknowledgments} 

I thank Dan Fabrycky and John Raymond for helpful discussions.  This work
was supported by in part by NASA grant NNX08AL43G, by FQXi, and by Harvard
University funds.

\end{document}